\documentstyle[floats,twocolumn,aps,psfig]{revtex}
\begin{document}

\draft

\title{Thickness evolution of coercivity in ultrathin magnetic films}
\author{R.A. Hyman$^1$, A. Zangwill$^1$, and M.D. Stiles$^2$}

\address{$^1$School of Physics, Georgia Institute of Technology, Atlanta, GA
30332-0430}
\address{$^2$Electron Physics Group, National Institute of Standards and
Technology, Gaithersburg, MD 20899}
\date{\today}
\maketitle

\begin{abstract}
The thickness evolution of in-plane magnetization reversal in
ultrathin films is studied with a theoretical model that takes account
of surface roughness typical of epitaxial growth. Guided by N\'{e}el's
model, step edge sites of monolayer-height islands are assigned a
two-fold anisotropy in addition to a four-fold anisotropy at all
sites. Coercivity is found to depend essentially on both the film
thickness and the partial coverage of the topmost layer. Its
qualitative features are determined primarily by sample geometry and
the size of the step anisotropy compared to the domain wall energy.
Magnetostatic interactions change the results quantitatively, but not
qualitatively.  Their effect is understood by comparing calculations
with and without their inclusion.
\end{abstract}


The potential for novel physics and exciting applications has
motivated many studies of ultrathin magnetic films \cite{HB}. To
explore new physics, it is usual to focus on simple model systems
and equilibrium properties such as exchange, anisotropy, and the
thermodynamic phase diagram in the space of temperature and
thickness. To exploit new applications, it is typical to study more
complex systems and non-equilibrium properties such as hysteresis,
domain wall motion, and magnetotransport. Common to both is the
observation that variations in surface roughness and film
morphology often have a significant effect on measurement results.
This observation motivated the theoretical work reported below.

Magnetometry \cite{Grad} and the surface magneto-optic Kerr effect
(SMOKE) \cite{Bader} are widely used to probe magnetization
reversal and the origin of coercivity in ultrathin films. A typical
experiment reports the coercive field as a function of total
deposited material using a thickness scale calibrated in small
fractions of a monolayer. While it is generally appreciated that
the films in question exhibit surface roughness, the consequences
of this fact are not often addressed explicitly. An exception is a
theoretical argument presented a few years ago by Bruno {\it et al.}
\cite{Bruno}  Making simple assumptions regarding thickness
fluctuations and the nature of domain wall pinning in films with
{\it perpendicular} magnetization, he derived a coercive field $H_C
\propto t^{-5/2}$ where $t$ is the film thickness. Experimental tests
of this prediction for the Co/Pd(111)
\cite{CoPd} and Co/Pt(111) \cite{CoPt} systems have yielded
contradictory results.

In this paper, we study hysteresis and coercivity as a function of
thickness for a model ultrathin film with {\it in-plane}
magnetization and surface roughness typical of as-grown samples. No
universal law is found. Instead, the coercive field is found to
depend on both the film thickness and the partial coverage of the
topmost incomplete layer. Our conclusion is based on
zero-temperature numerical simulations of magnetization reversal
using a generalization of a classical spin model of an ultrathin
film introduced previously \cite{Moschel}.

\begin{figure}
{\hspace{1.5in}
\psfig{figure=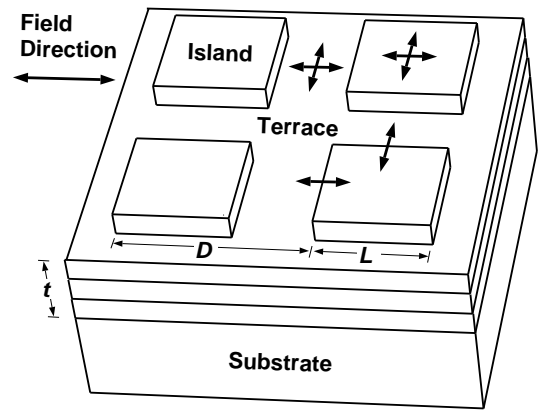,width=3.0in}}
\caption{Schematic view of the rough ultrathin film morphology used in
this work. The indicated island geometry is repeated periodically.
Arrows indicate local anisotropy axes for $K_4>0$ and $K_2<0$.}
\end{figure}

The model cubic film (Figure 1) sits atop a non-magnetic substrate
and takes the form of a periodic array of square, monolayer-height,
magnetic islands of side length $L$ and center-to-center separation
$D$ arranged on top of $1 \le t \le 5$ complete magnetic layers. To
model centered cubic lattices, the variable $t$ is measured in
units of $t_0=a/2$ where $a$ is the in-plane lattice constant.
Classical Heisenberg spins at each surface site $i$ point in the
direction $\hat{\bf S}_i$.
At each site in the surface, the spins are locked throughout the
thickness of the film giving a two-dimensional model.
This
two-dimensional model is appropriate if the film thickness is much
less than the exchange length.
Each spin is
subject to nearest-neighbor ferromagnetic exchange $J$, an out-of
plane surface anisotropy $K_Z>0$, a four-fold in-plane anisotropy
$K_4>0$, Zeeman energy from an external field $H$, and a two-fold
anisotropy $K_2$ at island perimeter sites only. A N\'{e}el model
analysis \cite{Gradmann} shows that the latter arises because step
edges break translational symmetry in the surface plane.  We also
take account of the fact that spins at step edges experience
reduced exchange compared to spins elsewhere on an island.

The magnetic energy is
\begin{eqnarray}
{ E_M} & = &  - \sum_{\langle i,j \rangle} J_{ij} \hat{{\bf S}}_i
\!
\cdot\! \hat{{\bf S}}_j
           - a^2K_2 \sum_{i \in step} (\hat{\bf S}_i \cdot \hat{\bf b}_i)^2
           \nonumber \\
    &   &  - 2a^2 K_4 \sum_{i}t_i\; [(\hat{S}^x_i)^4\! +\! (\hat{S}^y_i)^4]
           - \mu {\bf H}\! \cdot \! \sum_{i} t_i\hat{\bf S}_i \nonumber \\
    &   & + a^2K_Z \sum_i(\hat{S}^z_i)^2
\end{eqnarray}
where $t_i$ is the film height at site $i$ in units of $t_0$,
$\hat{\bf b}_i$ is a unit vector parallel to the local step edge,
$J_{ij}
= J\,{\rm min}[t_i,t_j]$, and $\mu =
\mu_0a^2t_0M_S$ where $\mu_0$ is the magnetic permeability in SI
units and
$M_S$ is the saturation magnetization. Note that, although $K_2>
0$ corresponds to an anisotropy axis parallel ($K_2> 0$ would be
perpendicular) to the step edge, for the highly
symmetrical island geometry studied here the
energy (1) is invariant to a change in the sign of $K_2$ and a
simultaneous rotation of the external field and all spins by
$90^{\circ}$ in plane.  Thus, the same reversal dynamics is expected
for both signs of $K_2$.
Typical values of the
material parameters, $J \sim 10^{-21} $ J and ${ K_4}
\sim 1
\times 10^{-3}$ mJ/m$^2$ (1 mJ/m$^2$=1 erg/cm$^2$), \cite{Heinrich}
imply a domain wall width
$W
\simeq 8
\sqrt{J/2K_4}
\simeq$ 600$a$. The numerical results reported below
all use the values $|K_2|=K_Z \sim 1$ mJ/m$^{\rm 2}$ \cite{Grad},
$a=0.3$ nm, and $\mu \sim 2\times 10^{-23}$ J/T in addition to
those quoted above for $J$ and $K_4$. The lengths $L$ and $D$ are
measured in spin block units of $10a \sim W/60$, a distance over
which no appreciable spin rotation occurs.

Beginning with a large positive value of ${\bf H} = H\hat{x}
\parallel [100]$, the local minimum of (1) was followed as the
field was reversed adiabatically by a combination of conjugate
gradient minimization and relaxational spin dynamics
\cite{Moschel}. The solid curves in Figure 2 illustrate the
variation of the computed coercive field $H_C$ with coverage
$\Theta = L^2/D^2$ when the incomplete magnetic layer sits atop
$t\!=\!1,3,5$ complete magnetic layers. Note that $H_C$ is
normalized to the Stoner-Wohlfarth value $H_{SW}=8a^2K_4/\mu$ where
easy-axis reversal occurs for a single domain system with four-fold
anisotropy \cite{SW}. Because $K_Z$ is large enough to keep the spins
in plane, the $t\!=\!1$ curve agrees
with our previous {\it XY}-model results \cite{Moschel}. They also agree
semi-quantitatively with SMOKE data reported by Buckley {\it et
al.} \cite{Buckley} for the Cu/Co/Cu(001) system. The rather
different $t\!=\!3$ and $t\!=\!5$ results can only reflect the
decreasing importance of the surface anisotropy, which scales like
$K_2^{*}
= K_2/t$ relative to other terms in the magnetic energy.
It is apparent from Figure 2 that
the coercivity depends not only on film thickness but also on the
partial coverage of the topmost incomplete layer. The reasons for
this are best understood by study of the corresponding
magnetization curves.

\begin{figure}
{\hspace{1.5in}
\psfig{figure=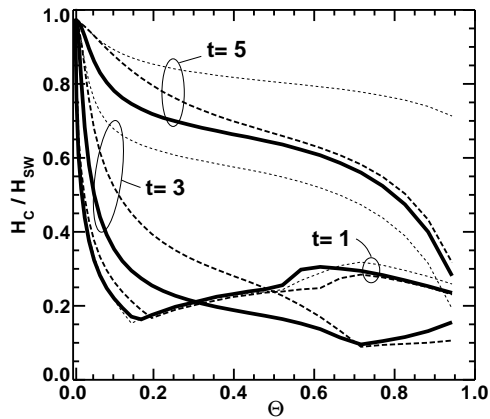,width=3.0in}}
\caption{Scaled coercive field as a function of $L/D$ when the
incomplete magnetic layer sits on $t=1,3,5$ complete magnetic
layers. $D=64$.  The solid curves give the results without
magnetostatics.  The heavy dashed curve and light dashed curve give
the results with magnetostatics for the cases of step anisotropy
parallel and perpendicular to the steps repsectively.}
\end{figure}

The hysteresis loops calculated during adiabatic cycling of the
field for $t\!=\!1,3,5$ and relatively {\it large} islands ($\Theta
\simeq {\textstyle{3 \over 4}}$) are shown in Figure 3. The
$t\!=\!1$ reversal scenario is understood \cite{Moschel,Hyman}.
Reversal initiates at those step edges where the anisotropy axis is
along the $y$ axis, (perpendicular to the saturated state spin
direction) because spins at those sites experience the greatest
torque. This first deviation from the positive saturated state
occurs at $H=H_N>0$, the nucleation field. The rotated spins form
small lens-shaped domains pinned at the islands edges within which,
near $H=0$, the spins are oriented nearly $90^{\circ}$ to the
original saturation direction. The domain walls depin from the
steps at the island edges at a (negative) field $|H|= H_S \sim
(W/L)H_{SW}$ \cite{notation} and the magnetization jumps to a
configuration where the $90^{\circ}$ state covers the islands
except near the complementary step edges where the anisotropy axis
is along the $x$ axis, i.e., where the two-fold axis locally pins
the spins parallel to the saturated state spin direction. Coherent
rotation of the island spins occurs as the external field is made
more negative (the coercive field occurs in this interval) until at
$|H|=H_T$ \cite{notation}, another magnetization jump occurs when
the $90^{\circ}$ single-domain state on the island terraces suffers
a Stoner-Wohlfarth instability to complete reversal (the
$180^{\circ}$ state). A final jump occurs when the external field
induces the complementary step-edge spins to flip from $0^{\circ}$ to
$180^{\circ}$ and the reversal is complete.

\begin{figure}
{\hspace{1.5in}
\psfig{figure=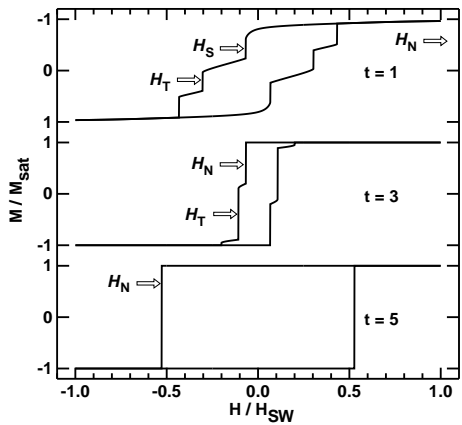,width=3.0in}}
\caption{Calculated film magnetization projected onto ${\bf H} \parallel
\hat{x}$ for films of thickness $t=1,3,5$ and large islands with $L=56$ and
$D=64$.}
\end{figure}

The $t\!=\!3$ loop in Figure 3 differs from the $t\!=\!1$ loop in
three important ways: (1) the nucleation field is negative rather
than positive and coincides with a jump in magnetization; (2) the
coercive field is smaller; and (3) there are two magnetization
jumps rather than three. Since nucleation occurs at two parallel
step edges per island, the origin of the change in sign of $H_N$
can be understood from the authors' recent analysis of magnetic
reversal on vicinal surfaces \cite{vicinal}. We showed there that,
for a single step (infinitely separated from other steps) on an
otherwise flat ultrathin magnetic film, reversal nucleates at
$H_N^{\infty} = H_{SW}({\cal K}^2-1)$ where ${\cal K}
= aK^{*}_2/2\sqrt{2JK_4}$ is a dimensionless measure of the
effective step anisotropy. Since ${\cal K} \sim 1$ for realistic
values of the magnetic parameters (as used here), it is not
surprising that $H_N$ can change from positive to negative as the
value of $K^{*}_2$ decreases. In fact, $|H_N|>H_S$ in this
instance, so nucleation is accompanied by an immediate jump to the
$90^{\circ}$ state on the islands.

Coherent rotation begins when the external field is made more
negative as in the $t\!=\!1$ case but the rotation is abruptly cut
off here by a jump of the island terrace spins to the $180^{\circ}$
reversed state. The corresponding jump occurred at a much larger
(negative) value of external field for the $t\!=\!1$ film. This
occurs because rotation of island terrace spins away from the
$90^{\circ}$ state costs exchange energy near every island edge due
to the two-fold anisotropy at each step. The reduction in $H_T$
seen in Figure 3 from $t\!=\!1$ to $t\!=\!3$ is thus expected
because the effective two-fold anisotropy is reduced for thicker
films.  It happens that $H_C=H_T$ in this case. Only a coherent
rotation of the complementary step spins is required to complete
reversal now because $K^{*}_2$ is here too small to strongly pin
those spins in the original saturation direction.

The $t\!=\!5$ hysteresis loop in Figure 3 is easily explained
because the formula quoted above for $H_N^{\infty}$ predicts an
even larger value for the nucleation field as $\tilde{K}_2$
continues to decrease. A simple square loop is found because
$|H_N|$ is greater than both $H_S$ and $H_T$. $H_C=H_N$ and the
entire film acts as a single domain particle.

\begin{figure}
{\hspace{1.5in}
\psfig{figure=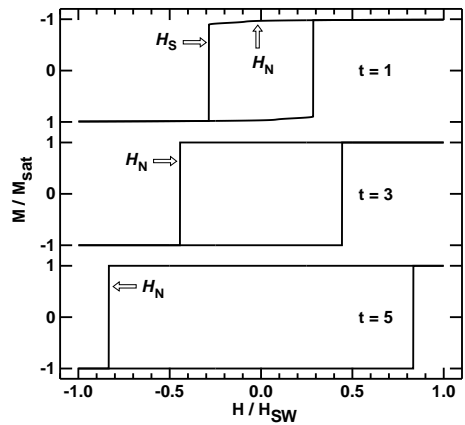,width=3.0in}}
\caption{Calculated film magnetization projected onto ${\bf H} \parallel
\hat{x}$ for films of thickness $t=1,3,5$ and small islands with $L=16$ and
$D=64$.}
\end{figure}

Figure 4 is the same as Figure 3 except that the magnetization
curves are shown for relatively {\it small} islands ($\Theta
\simeq 1/16$). The $t\!=\!1$ loop evidently differs drastically from
the corresponding loop for large islands. The nucleation field
$H_N>0$ but the much smaller island size $(L)$ drives the value of
$H_S \sim (W/L)H_{SW}$ so large that it exceeds $H_T$. A direct
jump to nearly complete reversal occurs at $H_C=H_S$. The change in
sign of $H_N$ explained above occurs between $t\!=\!1$ and
$t\!=\!3$ as before. But now the sequence $|H_N| > H_S > H_T$
guarantees that only a square loop will be found for $t\!=\!3$ in
Figure 4. The approach of $H_C=H_N$ to $H_{SW}$ observed for the
$t\!=\!5$ loop follows from the same argument based on
$H_N^{\infty}$ as was advanced earlier for the large island case.

Since $H_C=H_N$ for $t\ge 5$, the $t\!=\!5$ solid curve for $H_C$
in Figure 2 actually shows the island size dependence of $H_N$. The
results are comparable to the nucleation field dependence on step
separation for a periodic array of steps separated by a finite
distance\cite{vicinal}.  Those results, presented in Figure 7 of
Ref. \cite{vicinal}, show $H_N$ rapidly approaching zero (from
negative values) as the step spacing decreases through values
comparable to the distance between adjacent steps used in the
present work. Qualitatively, nucleation is easier when steps are
close together because the intervening spins tend to rotate also
and thus reduce the total exchange energy cost.

It has been argued that magnetostatics has a negligible effect on
the hysteresis of ultrathin films with in-plane magnetization
\cite{Arrott}. We checked this explicitly by adding a dipole-dipole
energy term to the magnetic energy (1). In the continuum limit,
this term can be written
\begin{equation}
E_D = -{\mu_0 \over 2}
\int d {\bf r} {\bf M \cdot H_D}
\end{equation}
where the dipolar field ${\bf H}_D$ is found by solving the Maxwell
equations $\nabla \cdot {\bf H}_D = - \nabla \cdot {\bf M}$ and
$\nabla \times {\bf H}_D =0$. In practice, these equations are
discretized and ${\bf M}_i = M_S \hat{\bf S}_i$ throughout the
thickness of the film at site $i$. In a system without step edges
any form of magnetization inhomogeneity is disfavored by
magnetostatics because the uniform state absolutely minimizes the
dipolar energy. In systems with step edges spin alignment parallel
to steps is favored to avoid exposed magnetic poles. This breaks
the symmetry noted earlier regarding the sign of $K_2$. Roughly
speaking, magnetostatics tends to enhance the step anisotropy when
$K_2<0$ (easy axis parallel to the steps) and weaken the step
anisotropy when $K_2>0$ (easy axis perpendicular to the steps). We
examined both cases.

The computational methodology used by us to evaluate the dipole
energy is similar to that described by Mansuripur
\cite{Mansuripur}. The curves with short (long) dashes in Figure 2
represent the coercive field including magnetostatics for the case
of $K_2>0$ ($K_2<0)$. Practically no effect is seen for $t\!=\!1$
as suggested in Ref. \cite{Arrott}. But the $t\!=\!5$ curves show
just the combination of effects expected qualitatively. Nucleation
at steps unavoidably generates magnetization inhomogeneities so a
larger driving force ($H_N$) is required to initiate reversal. This
effect is partly compensated when $K_2<0$ when the intrinsic step
anisotropy is parallel to the step. Conversely, nucleation is even
more retarded when magnetostatics works to further reduce
$K^{*}_2$.

When $H_C$ is determined by nucleation (all of the $t\!=\!5$ curve
and $t\!=\!3$ up to about $\Theta\! = \!3/4$), the systematics of
the magnetostatic results can be understood by noting that the
source of magnetostatic energy is $\nabla \cdot {\bf M}=
\partial_x M_x +
\partial_y M_y$. At saturation, $M_x$ is a constant and $M_y=0$. To
lowest order, $M_x$ remains constant at nucleation so only the
variations of $M_y$ in the $y$ direction contribute to $E_D$. We
have seen that the magnetization pattern at nucleation consists of
lens-shaped domains centered on those step edges where the local
anisotropy axis points in the $y$ direction \cite{unstable}. At the
smallest coverages when the islands are very small, $M_y =
M_y(x,y)$ is a function of both $x$ and $y$ so a finite
magnetostatic effect is expected. This  effect increases as the
island size increases initially because more step edge contributes
to nucleation. But for large enough island size, the magnetization
pattern at nucleation more nearly resembles the vicinal surface
case \cite{vicinal} and $M_y=M_y(y)$ when $K_2<0$ and $M_y=M_y(x)$
when $K_2>0$. The effect of magnetostatics thus increases in the
former case and disappears in the latter case as the islands grow
larger. This is the trend seen in Figure 2 except for the very
highest coverages of $t=3$ when $H_C$ is not determined by
nucleation \cite{Moschel,Hyman}.

Note also that magnetostatics has a weaker effect for $t=5$ than
for $t=3$. This is so because the thicker film experiences a
relatively weaker step anisotropy and behaves like a
Stoner-Wohlfarth particle. The magnetization is more uniform and
$\nabla
\cdot {\bf M}$ and $E_D$ are reduced accordingly.

In summary, we have used a classical spin model to study zero
temperature, in-plane, magnetization reversal in ultrathin films
with surface roughness characteristic of the epitaxial growth
process. The model includes four-fold anisotropy at all sites,
two-fold anisotropy at step edge sites, and magnetostatic
interactions. The coercive field was found to vary
non-monotonically (but explicably) as a function of both film
thickness and the partial coverage of the surface layer. We
conclude that measurements of the thickness dependence of the
coercive field and related hysteretic properties of ultrathin films
with in-plane magnetization should not be expected to exhibit
universal behavior. The results can depend sensitively on the film
morphology which, in turn, can depend sensitively on growth
conditions. Comparisons of nominally identical measurements on
nominally identical systems in ignorance of the film morphology is
unwarranted and can be seriously misleading.

R.\ A.\ H.\ acknowledges support from National Science Foundation Grant
No. DMR-9531115.  M.\ D.\ S.\ acknowledges useful conversations with
R.\ D.\ McMichael.


\begin{references}
\bibitem{notation}The symbols $H_S$ and $H_T$ are used to remind the
reader that the corresponding magnetization jumps are initiated by
spins on the islands steps and terraces, respectively. This usage
conforms to the notation used in Ref. \cite{vicinal}. The less
descriptive symbols $H_L$ and $H_{\perp}$ were used in Ref.
\cite{Moschel} and Ref. \cite{Hyman}.
\bibitem{unstable}The magnetization pattern at nucleation discussed here is
unstable if a jump in magnetization accompanies nucleation. But
this pattern is still used to determine the energy barrier to
nucleation. See Ref. \cite{vicinal} for more details.
\end{references}
\end{document}